\documentclass[twocolumn]{article}
\usepackage[left=0.5in, right=0.5in, top=0.7in, bottom=0.75in]{geometry}
\usepackage{graphicx}
\usepackage{float}
\usepackage{xcolor}
\usepackage{geometry}
\usepackage{multirow}
\usepackage{url}
\usepackage{xurl} 
\usepackage{bbold}
\usepackage{mathtools}
\usepackage{subfigure}

\usepackage[backend=biber,style=numeric-comp,sortcites,sorting=nty, natbib=true,hyperref,maxbibnames=99,url=false,eprint=false,doi=false,maxcitenames=1,firstinits=true]{biblatex}

\usepackage{hyperref}
\usepackage{amsmath}
\usepackage{algorithm}
\usepackage{algpseudocode}
\usepackage{xcolor}         
\usepackage{titlesec}
\titleformat*{\section}{\large\bfseries}
\titleformat*{\subsection}{\large\bfseries}
\titleformat*{\subsubsection}{\large\bfseries}

\hypersetup{
    pdftitle={Privacy-preserving Quantile Treatment Effect Estimation for Randomized Controlled Trials},
    pdfauthor={Yao et. al},
}

\begin{document}
\title{Privacy-preserving Quantile Treatment Effect Estimation for Randomized Controlled Trials}
\author{Leon Yao$^1$, Paul Yiming Li$^2$, Jiannan Lu$^2$}
\date{%
    $^1$MIT $^2$Apple\\[2ex]%
}
\maketitle
\begin{abstract}
In accordance with the principle of ``data minimization'', many internet companies are opting to record less data. However, this is often at odds with A/B testing efficacy. For experiments with units with multiple observations, one popular data minimizing technique is to aggregate data for each unit. However, exact quantile estimation requires the full observation-level data. In this paper, we develop a method for approximate Quantile Treatment Effect (QTE) analysis using histogram aggregation. In addition, we can also achieve formal privacy guarantees using differential privacy.

\end{abstract}

\section{Introduction}
A/B experimentation or Randomized Controlled Trials (RCTs) at internet companies often requires a massive amount of user data. Recently, there has been increased efforts among these companies to protect its user's data and the principle of ``data minimization'' is being implemented in fields like recommender systems and machine learning \cite{gdpr-recommender, gdpr-ml}.


There are also privacy-preserving methods for RCTs that comply with data minimization for estimating the Average Treatment Effect \cite{randomized-controlled-trials-apple, repeated-measurement-designs-netflix}. However, in some cases, Quantile Treatment Effects (QTE) are more preferred, as QTE is less skewed by outliers and is useful in analyzing effects at the tails of the distribution. Unfortunately, quantiles require a lot more data to estimate accurately. One way to aggregate user data to capture its distribution is to use a histogram. One benefit of histograms is that exact values are never recorded. 

In many cases, each unit may have repeated measurements or multiple observations. To account for this clustering, we will need user-level histograms. In this paper, we estimate difference-in-quantiles with cluster robust standard errors using histogrammed data. In addition to data minimization, we also provide formal privacy guarantees by using differential privacy (DP) \cite{dwork-dp}.

\section{Related Work}
Histograms have been widely used to estimate quantiles and is even included in popular software packages \cite{software-quantiles, ci_histogram, prometheus, hdrhist}. Some important considerations are the number of histogram bins, the size of each bin, and how quantiles are estimated within each bin. Without prior knowledge about the distribution of the data, common ways to determine bin widths include linear or log-linear spaced bins across some range. Using these histograms, we can easily determine which bin a specific quantile belongs to. The quantile is then estimated by linear interpolation, using the width of the bin \cite{prometheus}.

In the streaming or online computation literature, there are also many methods for approximating the quantile using data summaries or sketches \cite{MRL, GK, mergeable-summaries, dunning2019computing}. Some of these methods require storing a subset of the data, or using some strategic sampling \cite{MRL, GK, mergeable-summaries}. Unfortunately, storing exact user data is at odds with data minimization. Other methods might store summaries of the data, but it is unclear how we might estimate clustered standard errors \cite{dunning2019computing}.

Quantile estimation is also problem of growing interest in the DP community \cite{dwork-robust-statistics, pmlr-v139-gillenwater21a, pmlr-v162-kaplan22a}. Many also use the secure multi-party computation, federated learning or local differential privacy frameworks \cite{smpc-median, pillutla2022differentially, liu2023online}. However, none of these methods are adapted for the case of A/B experimentation. 


\section{Model}
\subsection{Delta Method for Clustered Data}
Let $X_1, X_2, \dots, X_n$ be n i.i.d.\ observations with mean $\mu$ and variance $\sigma^2$, where $\bar{X}$ is the sample mean. The central limit theorem states that as $n \rightarrow \infty$, then:

$$\sqrt{n}(\bar{X} - \mu) \rightarrow N(0, \sigma^2).$$

This asymptotic normality allows us to construct the $100(1-\alpha)\%$ confidence interval of $\mu$ as $\bar{X} \pm z_{\alpha/2} \sigma/\sqrt{n}$. The delta method allows us to generalize this to any function on $X$, such that when evaluated at $\theta$, $g'(\theta)$ exists and is non-zero: 

$$\sqrt{n} (g(X_n) - g(\theta)) \rightarrow N(0, \sigma^2 (g'(\theta))^2).$$

One application of the delta method is to estimate the mean of non-i.i.d.\ clustered observations. Let $i = 1, 2, \dots, K$, be the $i$th cluster with $N_i$ observations. We assume that within each cluster, the observations are i.i.d. Let $S_i = \sum_{j=1}^{N_i} X_{i,j}$. We can write the mean as:
$$\bar{X} = \frac{\sum_{i,j} X_{i,j}}{\sum_i N_i} = \frac{\sum_i S_i / K}{\sum_i N_i / K} = \bar{S}/\bar{N}.$$

Using the delta method on the ratio of two averages of i.i.d.\ observations, we can estimate the variance \cite{delta}:

\begin{equation}
\text{Var}(\bar{X}) \approx \frac{1}{K \bar{N}^2} \left[\text{Var}(S) - 2 \frac{\bar{S}}{\bar{N}} \text{Cov}(S, N) + \frac{\bar{S}^2}{\bar{N}^2} \text{Var}(N)\right].
\label{delta-clustering}
\end{equation}

Similarly, suppose $X_1, X_2, \dots, X_n$ are i.i.d.\ observations drawn from some distribution $f(X)$ with CDF $F(X)$. Let the order statistics $X_{(1)}, X_{(2)}, \dots X_{(n)}$, be the same observations in ascending order. We call the index of the order statistic the rank (i.e, the relative order in the sorted list). We can write the $p$-th sample quantile as $X_{(\lfloor np \rfloor)}$. Using the delta method \cite{castellaberger_stats}, we have :

\begin{equation*}
    \sqrt{n} \left(X_{(\lfloor np \rfloor)} - F^{-1}(p) \right) \rightarrow N \left(0, \frac{\sigma^2}{f\left(F^{-1}(p)\right)^2} \right).
\end{equation*}

However, this result is difficult to use as it requires estimating an unknown density function $f(X)$ and quantile $F^{-1}(p)$ \cite{liu2019largescale}. A more practical approach is the outer confidence interval method \cite{quantile_ci1, quantile_ci2}. Let $I_i = \mathbb{1}\{X_i \leq X_{(\lfloor np \rfloor)}\}$, such that $\sum_i I_i$ is the number of observations less than or equal to the sample quantile. This quantity is useful as $n \rightarrow \infty$, $\sqrt{n} (\bar{I} - p) \rightarrow N(0, \sigma^2)$, where $\sigma^2 = p(1-p)$.

The goal is to estimate the sample quantile and confidence interval in rank-space, then convert those to value-space. We then account for clustering using Eq. \ref{delta-clustering} \cite{delta}, by calculating a correction factor, $c$, for the confidence interval. The adjusted CI is then inverted to estimate the standard error of the sample quantile.

\begin{algorithm}
    \caption{Quantile Delta Method}
    \label{quantile-delta}
    \begin{algorithmic}[1]
        \State compute $L, U = n(p \pm z_{\alpha/2} \sqrt{p(1-p)/n})$
        \State fetch $X_{(\lfloor np \rfloor)}, X_{(L)}, X_{(R)}$
        \State compute $I_i = \mathbb{1}\{X_i \leq X_{(\lfloor np \rfloor)}\}$
        \State compute $\sigma_I = \text{Var}(\bar{I})$ using Eq. \ref{delta-clustering}
        \State compute correction factor $c = \sigma_I / \sqrt{p(1-p)}$
        \State compute $\sigma_p = \frac{c (X_U - X_L)}{2 z_{\alpha/2}}$
    \end{algorithmic}
\end{algorithm}

\subsection{Experimental Setup}
Under the potential outcome framework, we have units $i$ assigned to either treatment or control, $t_i \in \lbrace 0, 1 \rbrace$. Our observed outcome metric of interest for unit $i$ is denoted $Y_i = Y_i(t_i)$. We define the quantile function as $q(Y, p)$ for the $p$-th percentile. The quantile treatment effect (QTE) can then be written as:

$$\tau(p) = q(Y(1), p) - q(Y(0), p).$$

In our experiments, units are randomly assigned to treatment and control. Each unit can have multiple observations and within-unit observations are i.i.d. We can then estimate $q_{t} = \text{Var}(q(Y(t), p))$ using Algorithm \ref{quantile-delta}, and estimate $\text{Var}(\tau(p)) = \text{Var}(q(Y(1), p)) + \text{Var}(q(Y(0), p))$.

\subsection{Relative Lift}
In some cases, we may care about relative lift instead of absolute lift for our QTE. We define this as:
$$\tau(p) = q(Y(1), p) / q(Y(0), p) - 1.$$ 

Recall that Eq. \ref{delta-clustering} gives us the variance of the ratio of two averages of i.i.d quantities that are each asymptotically normal due to the central limit theorem. In this case, each average is just the quantile of treatment or control (i.e.\ $\bar{S} = q(Y(1), p)$ and $\bar{N} = q(Y(0), p)$ and are asymptotically normal. The variances $\text{Var}(q(Y, p))$ are estimated using Algorithm \ref{quantile-delta} and $\text{Cov}(S,N) = 0$ because treatment and control are independent:

\begin{equation}
    \text{Var}(\tau(p)) = \frac{1}{q_0^2} \left[ \text{Var}(q_1) + \frac{q_1^2}{q_0^2} \text{Var}(q_0) \right].
\end{equation}

\subsection{Histogram Method}
Instead of knowing the exact observations for every unit, we now consider aggregated data, where each unit will have it own histogram. We now need to adapt Algorithm \ref{quantile-delta} using histogram data. For a given quantile, $p$, we can still estimate the ranks, $\lfloor np \rfloor, L, U$ exactly. However, the quantities in value space, $X_{(\lfloor np \rfloor)}, X_{(L)}, X_{(R)}$, must be approximated. For a given rank, $r$, we can find which bin it belongs to and its relative rank, $k$, within the bin. Assume the bin contains $m$ observations and has left boundary $b_l$ and right boundary $b_r$. We can approximate $X_r$ by linearly interpolating using the bin width \cite{prometheus}:
\begin{equation}
    X_r \approx b_l + (b_r - b_l) \frac{k}{m}.
\end{equation}

We can also approximate $I_i = \mathbb{1}\{X_i \leq X_{(\lfloor np \rfloor)}\}$ by finding which bin $X_{(\lfloor np \rfloor)}$ belongs to and once again use linear interpolation to estimate the number of observations in this bin that are less than $X_{(\lfloor np \rfloor)}$. If we had a prior on the distribution of the observations we could also use some more sophisticated interpolation. 

We can now approximate the quantile using the delta method using histograms as shown in Algorithm \ref{quantile-delta-hist}.

\begin{algorithm}
    \caption{Quantile Delta Method using Histograms}
    \label{quantile-delta-hist}
    \begin{algorithmic}[1]
        \State compute $L, U = n(p \pm z_{\alpha/2} \sqrt{p(1-p)/n})$
        \State approx. $\tilde{X}_{(\lfloor np \rfloor)}, \tilde{X}_{(L)}, \tilde{X}_{(R)}$ with interpolation
        \State approx. $\tilde{I}= \mathbb{1}\{X_i \leq X_{(\lfloor np \rfloor)}\}$ with interpolation
        \State compute $\sigma_I = \text{Var}(\bar{\tilde{I}})$ using Eq. \ref{delta-clustering}
        \State compute correction factor $c = \sigma_I / \sqrt{p(1-p)}$
        \State compute $\sigma_p = \frac{c (X_U - X_L)}{2 z_{\alpha/2}}$
    \end{algorithmic}
\end{algorithm}

\subsubsection{Bin Widths}
Histograms are defined by the number of bins and the width of each bin. The more bins we use the better accuracy we have when estimating quantiles. However, we cannot have too many bins otherwise we will weaken the data minimization. Our goal is to select the number of bins and widths to fit our analysis goals.

Two simple methods are to use linearly or log-linearly spaced bins across the range of the observations. The problem with these binning strategies is that real data are often normally distributed or even power-law \cite{clauset-power-law}, which means that a few bins contain a disproportionate number of observations. This will lead to increased error in our quantile estimates.

Ideally, we would a priori know the exact distribution of the observations in our experiment. We would then use the exact quantiles of the distribution (denoted ``PX'' for the Xth quantile) as the bin boundaries. Depending on the granularity of the desired quantiles, we would need an inversely proportionate amount of bins. For example if we wish to estimate the quantiles P1, P2, $\dots$, P99, then we need at least 100 bins. However, realistically, the best we can do is use some historical distribution of the metric and have the bins boundaries be quantiles of this prior distribution.

\subsection{Differential Privacy}
A function $f$, is $\epsilon$-differentially private if for all ``neighboring'' databases $D_1$ and $D_2$, which differ in only one entry, for some outcome $O$:
\begin{equation}
    \mathbb{P}[f(D_1)=O] \leq e^\epsilon \cdot \mathbb{P}[f(D_2)=O].
\end{equation}

We can make the function $f$ $\epsilon$-differentially private by adding noise to the output drawn from the Laplace($\Delta/\epsilon$) distribution. We define $\Delta$ as the $l_1$-sensitivity of $f$ as:

\begin{equation}
    \Delta = \max_{D_1, D_2} ||f(D_1) - f(D_2)||_1.
\end{equation}

In other words, the sensitivity is the most the function can change by adding/removing a single entry. In the case of histograms, each bin is a separate function that counts how many observations fall between its boundaries. The sensitivity is 1, since a single observation can at most change the value of the bin by 1. We can then make our observations $\epsilon$-DP by adding noise from the Laplace distribution to each bin. Because the histogram has only integer values and cannot go below 0, we use the discrete version of the Laplace distribution, known as the geometric distribution \cite{ghosh2012universally} and truncate negative values to 0.

\section{Data}
\vspace{-3mm}
\begin{table}[H]
\centering
\begin{tabular}{l|ll}
Experiment \# & \# Units  & \# Observations \\ \hline
1             & 6,000,000 & 23,000,000      \\
2             & 100,000   & 1,000,000       \\
3             & 50,000    & 600,000        
\end{tabular}
\caption{Number of units and observations for 3 real-world RCTs we use to evaluate our methodology.} \label{table:exp-data}
\end{table}

We evaluate our methods on three real-world RCTs from a large internet company with stats shown in Table \ref{table:exp-data}. For each observation we look at three key performance outcome metrics with contrasting distributions. For each metric, we also clip the values to a lower and upper threshold, such that all values outside the bounds are set to the threshold values. We determine bin boundaries using historical data; for each metric we find its distribution across all units over a large time span. We then use the distribution's exact quantiles as the bin boundaries. When we say ``100 bins'', we use 100 equally spaced quantiles across the historical distribution (i.e P1, P2, $\dots$, P99). 

\section{Evaluation Metrics}
\subsection{Full-data baseline comparison}
We can measure the error of the histogram delta method by comparing it to the delta method approach on the full original dataset, where we have non-aggregated observations for each unit. Specifically, we can measure the relative error of the quantile point estimate and standard error. In practice, we do not need to estimate arbitrary quantiles. Many applications care only about a few quantiles such as P50, P95 and P99. We can then measure the error for various values of bin count across the three bin boundary methods (linear, log-linear, and quantiles of historical data). 

\subsection{A/A permutation test}
Additionally, we can evaluate how histogram data affects the false positive rate of our QTE estimator using the A/A permutation test. Under the null hypothesis of no treatment effect, we repeatedly permute the treatment assignments of our experiment and estimate a p-value. If the distribution of p-values are uniform (tested using a Kolmogorov-Smirnov test), then the QTE has correct false positive rate for every significance level \cite{castellaberger_stats}. This p-value uniformity test may be too strict, as we may care only about the coverage of a few common significance levels such as 0.01 and 0.05. Using these A/A permutations, we can also measure whether their confidence intervals cover 0 at the correct rate. 
 \section{Results}
We run our method on all three experiments and metrics, but we mainly show the results for the same metric on experiment 2 for consistency. The results generalize well across experiments.

\begin{figure}[!ht]
    \centering
    \subfigure[Historical Quantile Comparison]{
        \includegraphics[width=.8\columnwidth]{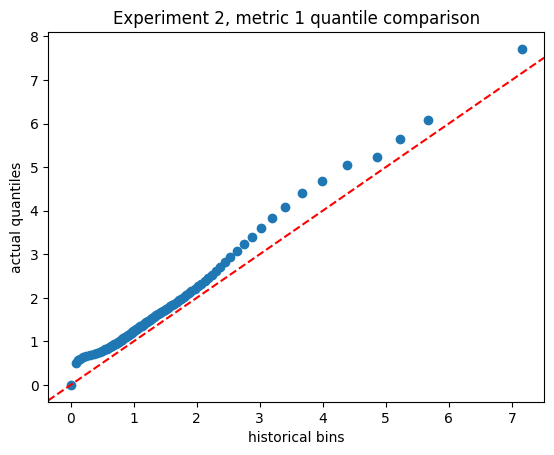}\label{fig:quantile-scatter-exp2-m1}
    }
    \subfigure[Bin Boundaries Comparison]{
        \includegraphics[width=.8\columnwidth]{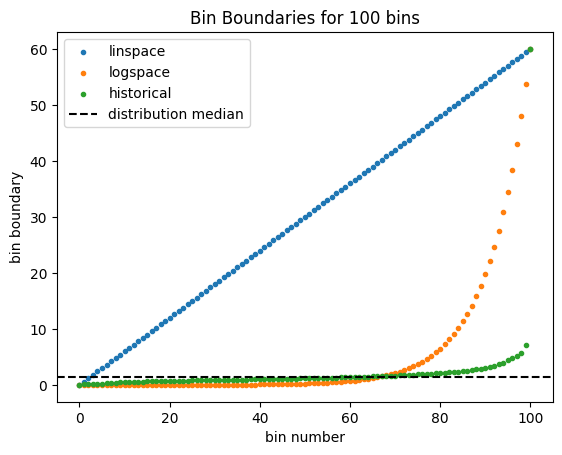}\label{fig:100-bin-comp}
    }
    \caption{On top, we plot a scatter-plot of the quantiles in 1\% intervals for the historical distribution and the actual distribution of a metric for experiment 2. On bottom, we compare the bin boundaries for the 3 binning strategies.} \label{fig:quantile-scatter}
\end{figure}
We first wish to evaluate how well our binning strategies perform. For each experiment's distribution for each metric, we can compare the actual vs historical bin boundaries, as shown in Fig. \ref{fig:quantile-scatter-exp2-m1}. We see that there is some mismatch in certain parts of the distribution. In Fig. \ref{fig:100-bin-comp}, we see that for this particular metric, the range of the distribution goes from 0 to 60. Linearly spaced bins (``linspace'') has too many bins for the right tail of the distribution, as the distribution median falls in bin 3. Log-linear bins (``logspace'') appears to do better, however, between p25 and p75, there are still only 7 bins. Historical bins, in comparison, have 39 bins in this range. The distribution of the bins have significant effects on how well a particular binning strategy can perform across various quantiles.

\begin{figure}[!ht]
    \centering
    \subfigure[P50 QTE error]{
        \includegraphics[width=.8\columnwidth]{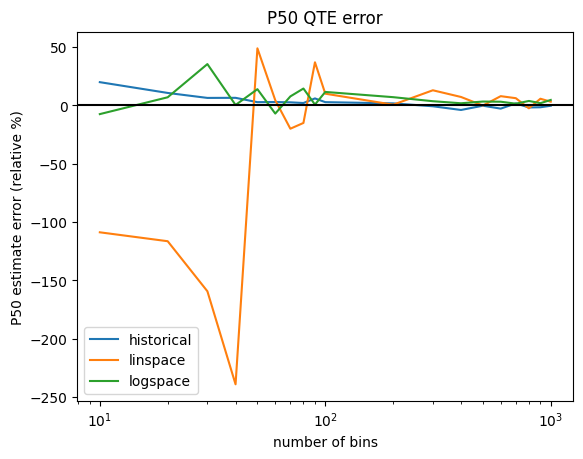}\label{fig:p50-error}
    }
    \subfigure[P99 QTE error]{
        \includegraphics[width=.8\columnwidth]{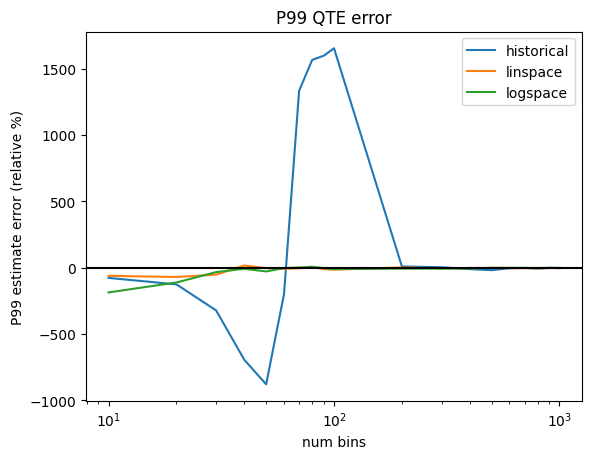}\label{fig:p99-errpr}
    }
    \caption{For the 3 binning strategies, we compare the P50 and P99 QTE estimate for various numbers of bins. We plot the relative error compared to the full-data baseline.} \label{fig:quantile-error}
\end{figure}

In Fig. \ref{fig:quantile-error}, we compare the relative error of the QTE estimate for the three binning strategies by varying the number of bins. As expected, the P50 estimate using historical bins has the least error, since it has the highest granularity in bin count near that quantile. However, for the P99 estimate with 100 historical bins, only 1 bin covers that quantile, which causes huge errors. Meanwhile, 100 linspace bins does extremely well for P99, since 88 bins are near that quantile. As we increase the number of bins, the granularity for the historical bins at P99 increases and the error decreases dramatically.

\begin{table*}[!ht]
\centering
\begin{tabular}{c|cccccc}
\# bins  & P50    & 95\% CI width & P95    & 95\% CI width & P99     & 95\% CI width \\ \hline
Baseline & 0.0144 & 0.024         & -0.178 & 0.11          & 0.152   & 0.44          \\ \hline
20       & +9.6\% & +3.3\%        & +437\% & +3400\%       & -123\%  & +764\%        \\
50       & +2.2\% & -1.5\%        & +20\%  & +20\%         & -894\%  & +1090\%       \\
100      & +0.5\% & +0.8\%        & +2.3\% & -4.6\%        & +1660\% & +1680\%       \\
200      & +2.3\% & -0.4\%        & -8.6\% & -6.0\%        & +8.1\%  & +23.7\%       \\
500      & -0.8\% & +6.1\%        & -2.3\% & -3.8\%        & -16.5\% & -0.6\%        \\
1000     & -2.1\% & +4.2\%        & +4.8\% & +0.7\%        & -0.4\%  & +7.6\%       
\end{tabular}
\caption{We compare the QTE point estimate and 95\% CI width using historical bins for P50, P95 and P99 for experiment 2. The first row shows the values for the full-data baseline. Then we show the relative values for the histogram approximation across various number of bins.} \label{table:baseline-comp}
\end{table*}

Now, we compare the histogram method using the historical method with the full-data baseline. In Table \ref{table:baseline-comp}, we report the QTE point estimate and 95\% CI width for P50, P95 and P99 across various numbers of bins. The values are shown as a percentage change from the baseline. We can interpret this as how much bias and how much the CI widens or shrinks when using the histogram method. In general, adding more bins improves the point estimate and 95\% CI estimate. The more granular the quantile, the more bins that are needed to get an accurate estimate. Additionally, there is more error near the tails of the distribution because of the clipping. For example, for P99, using 20, 50, or even 100 bins do not give close estimates. A minimum of 1000 bins is needed for reasonable P99 results. 

\begin{table}[!ht]
\setlength\tabcolsep{4pt}
\begin{tabular}{c|cccc}
Quantile             & Metric & KS p-value & 95\% Cover & 99\% Cover \\ \hline
\multirow{3}{*}{P50} & 1      & 5.6E-08    & 95.4\%     & 99.4\%     \\
                     & 2      & 0.28       & 94.7\%     & 99.0\%     \\
                     & 3      & 2.6E-14    & 95.6\%     & 99.4\%     \\ \hline
\multirow{3}{*}{P95} & 1      & 0.019      & 94.7\%     & 98.9\%     \\
                     & 2      & 8.2E-03    & 95.8\%     & 99.2\%     \\
                     & 3      & 0.41       & 95.6\%     & 99.2\%     \\ \hline
\multirow{3}{*}{P99} & 1      & 9.1E-06    & 95.4\%     & 99.0\%     \\
                     & 2      & 0.34       & 94.9\%     & 98.9\%     \\
                     & 3      & 4.5E-03    & 94.5\%     & 99.0\%    
\end{tabular}
\caption{A/A Test results for histograms with 1,000 bins over 10,000 randomizations for experiment 2. We report the KS uniformity test p-value (p $\geq$ 0.05 means it passes) and the 95\% and 99\% CIs cover \%.} \label{fig:aa-test}
\end{table} 

Finally, we evaluate the false positive rate using the A/A permutation test on 10,000 randomization draws. In Table \ref{fig:aa-test}, we see using histograms with 1,000 bins has about half the cases pass the p-value uniformity test. Even in cases when it does not pass, the 95\% and 99\% CI cover \% has small error, which suggests that our QTE estimator has appropriate false positive rate.

\subsection{Differential Privacy}

\begin{figure}[!ht]
    \centering
    \subfigure[QTE]{
        \includegraphics[width=.8\columnwidth]{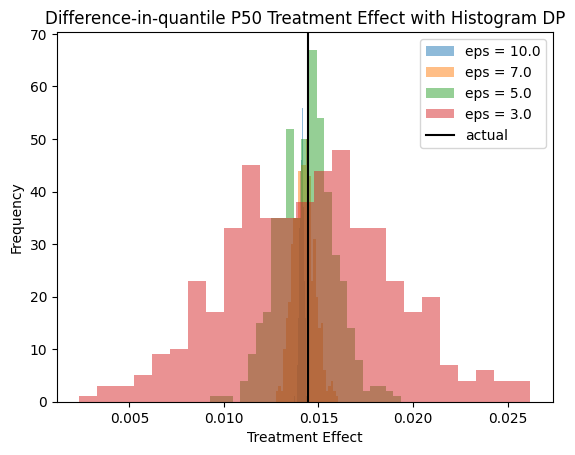}\label{fig:quantile-te-dp}
    }
    \subfigure[Standard Error]{
        \includegraphics[width=.8\columnwidth]{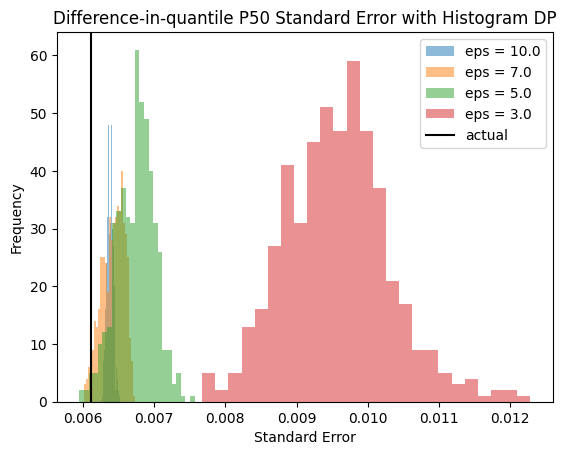}\label{fig:quantile-se-dp}
    }
    \caption{We draw 500 randomizations of the DP noise and apply it to the histogram data for experiment 2. We then estimate the P50 QTE and standard error.} \label{fig:quantile-dp}
\end{figure}

We also simulate the effects of Differential Privacy  on the histogram method. Recall that we can make the histogram data differentially private by adding noise drawn from the geometric distribution. The amount of privacy we achieve is tuned by the $\epsilon$ parameter. The smaller the $\epsilon$, the more noise is added, and the more privacy is guaranteed. For various values of $\epsilon$, we simulate the effects of DP by repeatedly taking draws from the geometric distribution, applying it to the histogram data, and estimating the QTE and its standard error using the histogram method. These $\epsilon$ values are in-line with many real-world applications of DP \citep{desfontainesblog20211001}. In Fig. \ref{fig:quantile-dp}, we show the distribution of results after 500 draws for each $\epsilon$. We find that as $\epsilon$ decreases and more noise is added, the variance of the QTE increases and there is bias on its standard error. 

\section{Practical Considerations}
Many large internet companies with extensive experimentation platforms record hundreds of metrics per experiment. In many cases experiments are run to test new features, meaning that the distribution of the observation should be different than the historical distribution. It may be possible to adaptively learn proper bin widths using early data. There may even be benefits for the precision of our method by storing segments of data with different bin widths. However, the amount of data storage required quickly adds up for these more advanced methods. In many cases, units are a part of multiple experiments and have multiple observation metrics. This would mean each unit would store a separate histogram for each experiment and metric. In practice, experiments should use the same bin boundaries derived from the same historical distributions.

As we see from our results, using historical distribution quantiles as bin boundaries works well for the majority of quantiles, but may do poorly at the tails of the distribution. The static binning strategies only do well for the quantiles where the bins are more granular. It is easy to arbitrarily decrease the error of our estimates by adding more bins, however, this adds more privacy concerns. 

One way to balance the utility vs privacy trade-off is to select the quantiles of the historical distribution that matters most for our analysis. The results shown use equally spaced quantiles across the entire distribution, which assumes that we equally care about all quantile levels. However, as we discussed previously, we may only care about a certain few, like P50, P95 and P99. We can instead add more granular bins around these quantiles of interest.

\section{Conclusion}

Using histogram aggregated data, we show that we can approximate quantile treatment effects and their confidence intervals accurately. Although there may be more complicated methods that can better approximate these values (more advanced interpolation, better quantile sketches, adaptive bin boundaries), they offer more difficult implementation and storage challenges. The simplicity of histograms allows for efficient and understandable data storage, and also offers an easy way to add differential privacy for formal privacy guarantees. Depending on additional data curation restrictions, local differential privacy can also be easily implemented in this framework \cite{ldp-hist}.

\printbibliography

\end{document}